\documentstyle[12pt]{article}
\makeatletter\newdimen\normalarrayskip            
\newdimen\minarrayskip                 
\normalarrayskip\baselineskip
\minarrayskip\jot
\newif\ifold             \oldtrue            
\def\arraymode{\ifold\relax\else\displaystyle\fi}
\def\eqnumphantom{\phantom{(\theequation)}}
\def\@arrayskip{\ifold\baselineskip\z@\lineskip\z@
     \else
     \baselineskip\minarrayskip\lineskip2\minarrayskip\fi}
\def\@arrayclassz{\ifcase \@lastchclass \@acolampacol \or
\@ampacol \or \or \or \@addamp \or
   \@acolampacol \or \@firstampfalse \@acol \fi
\edef\@preamble{\@preamble
  \ifcase \@chnum
     \hfil$\relax\arraymode\@sharp$\hfil
     \or $\relax\arraymode\@sharp$\hfil
     \or \hfil$\relax\arraymode\@sharp$\fi}}
\def\@array[#1]#2{\setbox\@arstrutbox=\hbox{\vrule
     height\arraystretch \ht\strutbox
     depth\arraystretch \dp\strutbox
     width\z@}\@mkpream{#2}\edef\@preamble{\halign \noexpand\@halignto
\bgroup \tabskip\z@ \@arstrut \@preamble \tabskip\z@ \cr}%
\let\@startpbox\@@startpbox \let\@endpbox\@@endpbox
  \if #1t\vtop \else \if#1b\vbox \else \vcenter \fi\fi
  \bgroup \let\par\relax
  \let\@sharp##\let\protect\relax
  \@arrayskip\@preamble}
\def\eqnarray{\stepcounter{equation}%
              \let\@currentlabel=\theequation
              \global\@eqnswtrue
              \global\@eqcnt\z@
              \tabskip\@centering
              \let\\=\@eqncr
              $$%
 \halign to \displaywidth\bgroup
    \eqnumphantom\@eqnsel\hskip\@centering
    $\displaystyle \tabskip\z@ {##}$%
    &\global\@eqcnt\@ne \hskip 2\arraycolsep
         \hfil$\arraymode{##}$\hfil
    &\global\@eqcnt\tw@ \hskip 2\arraycolsep
         $\displaystyle\tabskip\z@{##}$\hfil
         \tabskip\@centering
    &{##}\tabskip\z@\cr}
\makeatother

\def\beq{\begin{equation}}
\def\eeq{\end{equation}}
\def\bea{\begin{eqnarray}}
\def\eea{\end{eqnarray}}

\begin{document}
\begin{titlepage}
\begin{center}
\hfill UUITP 11/94\\
\hfill hep-th/9411148

\vspace{0.1in}{\Large\bf Some properties of the linearized model \\
of the (super)$p$-brane }\\[.4in] {\large P.Demkin}\footnote{On leave
from Department of Physics, Vilnius University, Saul\.{e}tekio al.9,
2054, Vilnius, Lithuania}\\
\bigskip {\it
Institute of Theoretical Physics \\ Uppsala University\\ Box 803,
S-75108\\ Uppsala, Sweden\\ paul@rhea.teorfys.uu.se}

\end{center}
\bigskip \bigskip

\begin{abstract}
Some general properties of the relativistic $p$-dimensional surface
imbedded into $D$-dimensional spacetime and its reduction to the
sim\-plest case of the quadratic Lagrangian (the linearized model) are
considered. The solutions of the equations of motion of the linearized
model for the $p$-brane with arbitrary topology and massless
eigenstates, as well as with critical dimension after quantization are
presented. Some generalizations for the supermembrane are discussed.

PACS Nos.: 03.70, 11.17.
\end{abstract}

\end{titlepage}
\newpage
\section{Introduction}

Nowadays not only one-dimensional relativistic objects - strings, but
also the objects of higher dimension - $p$-dimensional (super)$p$-branes
are suggested as substantial physical and mathematical objects. As for
their properties, much less about those of $p$-branes is known this far
[1-7].

The necessity to consider multidimensional objects with more than one
space dimensions arises in various parts of the field theory. In
particular, we may try to consider the (super)$p$-brane theory as
fundamental, like the (super)string theory ($p=1$) \cite{B1}, as well as
an effective model of supergravity, as shown in \cite{Hu}.
A possible correlation between ordinary and rigid (super)$p$-branes and,
in particular, the correlation between the rigid string and the ordinary
membrane at $p$=2 has been considered in \cite{Li,P1}.
The culculation of the static potential for the $p$-brane compactified on the
 space-times of the various forms has been considered in \cite{BO2, BZ2}

For the supermembrane ($p=2$), action is a direct multidimensional
generalization of the string action \cite{B1}:
\begin{equation}
S=-\frac{T}{2} \int
d^{3}\xi[\sqrt{h}h^{ij}\Pi^{a}_{i}\Pi^{b}_{j}\eta_{ab} - \sqrt{h}+
2\varepsilon^{ijk}\Pi^{A}_{i}\Pi^{B}_{j}\Pi^{C}_{k}B_{CBA}],
\end{equation}
where $T$ is the parameter of tension with the dimension \-
$(mass)^{(p+1)}$ \- or $(length)^{-(p+1)}$, $\xi^{i}$ $(i=0,1,..., p)$
are the worldvolume coordinates, $h_{ij}$ is the metric of the
worldvolume, $h=-$det$(h_{ij})$, $\eta_{ab}$ is the Minkowski spacetime
metric, and $\Pi^{A}_{i}=\partial_{i}Z^{M}E^{A}_{M},\; A=a,\alpha;\;
M=\mu, \dot{\alpha}$. Here, $Z^{M}$ are the coordinates of the
$D$-dimensional curved superspace, and $E^{A}_{M}$ is the
supervielbein. The 3-form $B=\frac{1}{6}E^{A}E^{B}E^{C}B_{CBA},\;
E^{A}=dZ^{M}E^{A}_{M}$ is the potential for the closed 4-form $H=dB$.

The action (1) is invariant respecting the global $D$-dimensional
Poincar\'{e} transformations, as well as it is invariant respecting
local parametrizations of the worldvolume with the parameters
$\eta^{i}(\xi)$:
\begin{equation}
\delta Z^{M} = \eta^{i}(\xi)\partial_{i}Z^{M},
\quad \delta h_{ij} =
\eta^{k}\partial_{k}h_{ij} + 2\partial_{(i}\eta^{k}h_{j)k} \;.
\end{equation}
It is also invariant under local fermionic ``$k$-transformations'':
\begin{eqnarray}
\delta Z^{M}E^{a}_{M} = 0,\qquad \qquad \qquad \qquad \\
\delta Z^{M}E^{\alpha}_{M} = (1+\Gamma)^{\alpha}_{\beta}k^{\beta},
\qquad \qquad  \qquad \\
\delta(\sqrt{h}h^{ij}) = -2i(1+\Gamma)^{\alpha}_{\beta}k^{\beta}
(\Gamma_{ab})_{\alpha\gamma}
\Pi^{\gamma}_{n}h^{n(i}
\varepsilon^{j)kl}\Pi^{a}_{k}\Pi^{b}_{l}- \nonumber \\
-\frac{2i}{3\sqrt{h}}
k^{\alpha}(\Gamma_{c})_{\alpha\beta}
\Pi^{\beta}_{k}\Pi^{c}_{l}h^{kl}\varepsilon^{mn(i}
\varepsilon^{j)pq}\times
\nonumber\\ \times(\Pi^{a}_{m}\Pi_{pa}\Pi^{b}_{n}\Pi_{qb} +
\Pi ^{a}_{m}\Pi_{pa}h_{nq}  + h_{mp}h_{nq}),
\end{eqnarray}
with an anticommuting spacetime spinor $k^{\alpha}(\xi)$, and the matrix
$\Gamma$ defined by
\begin{equation}
\Gamma=\frac{1}{6\sqrt{h}}\varepsilon^{ijk}
\Pi^{a}_{i}\Pi{b}_{j}\Pi^{c}_{k}\Gamma_{abc}\;.
\end{equation}

Unlike the two-dimensional string action, the action (1) at $p\neq 1$ is
not invariant respecting local conformal transformations with the
parameter $\Lambda (\xi)$:
\begin{eqnarray}
\delta Z^{M}=0;\\
\delta h^{\alpha \beta}=\Lambda (\xi)h^{\alpha \beta}\; .
\end{eqnarray}

Varying the initial action leads to essentially non-linear field
equations
\begin{eqnarray}
\partial_{i}(\sqrt{h}h^{ij}\Pi^{a}_{j})+
\sqrt{h}h^{ij}\Pi^{b}_{j}\Pi^{C}_{i}\Omega_{Cb}^{a} + i\varepsilon^{ijk}
\Pi_{ib}(\Pi^{\alpha}_{j}
\Gamma^{ab}_{\alpha\beta}\Pi^{\beta}_{k}) +\nonumber\\
+\varepsilon_{ijk} \Pi^{b}_{i}\Pi^{c}_{j}\Pi^{d}_{k}H^{a}_{bcd} = 0,\\
\ [ (1-\Gamma )h^{ij}\Pi^{\mu}_{i}\Gamma_{\mu}] ^{\alpha}_{\beta}
\Pi^{\beta}_{j} = 0,
\end{eqnarray}
where $\Omega_{B}^{A}$ is the 1-form connection in the $D$-dimensional
curved superspace, and to the "embedding" equation
\begin{equation}
h_{ij} = \Pi^{a}_{i}\Pi^{b}_{j}\eta_{ab}\;,
\end{equation}
which remains non-linear at any gauge. Their solution is known for
certain simplest cases \cite{B2}.

For open membranes, or for the existing open dimensions, at $\sigma
_{i}=\sigma _{i}^{a}, \sigma _{i}=\sigma_{i}^{b}$ the border condition
is observed on the coordinates $Z^{M}(\xi )$:
\begin{equation}
\int d^{3}\xi \partial _{i}(\delta Z^{a}\sqrt{h}h^{ij}\Pi _{ja}+
3\varepsilon^{ijk}\delta Z^{A}\Pi ^{B}_{j}\Pi ^{C}_{k}B_{CBA})=0,
\end{equation}
where $h_{ij}$ is given by equation (11).

Any new solution of the equations of motion (9) and (10) describing the
motion of a multidimensional relativistic object, on one hand, is of
interest in itself, and on the other hand, it serves as a starting point
for semiclassical quantization, when the minor variations respecting the
known classical solution are investigated.

We have considered a mathematically simpler case at $p$=2. M.Duff in
\cite{D2} presents a $p$-dimensional generalization of the supermembrane
action, which has similar properties.

In the case when a complicated, non-linear dynamic system is
investigated, it seems reasonable to start from its linearized
model. This work aims to investigate a special type of action
corresponding to a linerized model of the relativistic
(super)$p$-brane. Such approach is possible in all cases when the
(super)$p$-brane model appears.

\section{A linearized model of the bosonic $p$-brane}

Let us consider as a less complicated the case of the bosonic
relativistic $p$-brane. This means that we are considering the action
\begin{equation}
S = -T\int
d^{p+1}\xi|det(\partial_{\alpha}X^{\mu}\partial_{\beta}X^{\nu}g_{\mu
\nu})|^ {\frac{1}{2}} ,
\end{equation}
where $\xi=(\tau,\sigma_{1},\ldots,\sigma_{p})$, \quad
$\xi_{\alpha}\in[\xi^{a}_{\alpha},\xi^{b}_{\alpha}]$, \quad
$X^{\mu}=X^{\mu}(\tau,\sigma_{1},\ldots,\sigma_{p}) , \;
\mu=0,\ldots,D-1$, where $D$ is the dimension of the Minkowski spacetime
with the metric $g_{\mu\nu}; \; \alpha=0,\ldots,p$, where $p$ is the
space dimension of $p$-brane.

The equation of motion
\begin{equation}
\partial_{\alpha}(\sqrt{h}h^{\alpha \beta}\partial_{\beta}X^{\mu})=0,
\end{equation}
resulting from (13), in the case the border conditions are taken into
account, may be obtained from the classically equivalent action

\begin{equation}
S = -\frac{T}{2} \int d^{p+1}\xi \sqrt{h} [h^{\alpha \beta}
\partial_{\alpha} X^{\mu}\partial_{\beta}X^{\nu}g_{\mu \nu} - (p-1)]\;,
\end{equation}
where an auxiliary metric $h_{\alpha \beta}$ on the worldvolume of the
membrane is introduced. The actions (13) and (15) to be equivalent, the
metric $h_{\alpha \beta}$ must obeys the imbedding condition:
\begin{equation}
h_{\alpha \beta } =\partial _{\alpha }X^{\mu }\partial _{\beta}X^{\nu}
\eta_{\mu \nu }\;,
\end{equation}
like the embedding condition (11) in the supersymmetric case.

Besides, we must check if the constraint conditions $p+1$ are observed:
\begin{eqnarray}
P^{\mu}_{\tau}X_{\mu ;i}=0, \qquad P^{2}+T^{2}deth_{ij}=0,
\end{eqnarray}
where $P^{\mu}_{\tau}=\delta \cal L / \delta $\.{X}$^{\mu}, \quad 1\leq
i,j\leq p$.

We cannot quantize action (15) at $p > 1$, but we can introduce a
certain simplification. Let $Y^{\mu}$ be a variation respecting the
classical solution $X^{\mu}_0$:
\begin{equation}
X^{\mu}=X^{\mu}_0 +\varepsilon Y^{\mu}\; .
\end{equation}

Then the equation of motion (14) turns into
\begin{equation}
\partial_{\alpha}P^{\mu \alpha}= \partial_{\alpha}
P_0^{\mu \alpha}+
\varepsilon\partial_{\alpha}C^{\mu \alpha}+o(\varepsilon)=
0\; .
\end{equation}

The requirement of the $X^{\mu}$-solution of the equation of motion
being the first order in $\varepsilon$ leads to the equation
$\partial_{\alpha}C^{\mu \alpha}=0$:
\begin{equation}
\partial_{\alpha} \frac{\partial A}{\partial X_{\mu;\alpha}} +{3 \over 2h^{0}}
\sum_{\alpha=0}^{p}
\partial_{\alpha}(A{\partial h^0 \over \partial \dot{X}_{0\mu }})=0,
\end{equation}
where
$A=\sum_{i,j=0}^{p}\partial_{i}X_{0}^{\mu}\partial_{j}Y_{\mu}
\bar{h}^{0}_{ij},\quad
h^{0}_{ij} =\partial_{i}X^{\mu}_{0}\partial_{j}X_{0\mu}, \quad
h^{0}=deth^{0}_{ij}\;$.

The exact expression for the equation of motion (20) depends on the
solution $X^{\mu}_{0}(\xi)$. For instance we may consider special tipe of the
solution with one or few compactified dimensions. The solution for the toroidal
membrane on the spacetime with the topology $R^{D-2}\times S^{1}\times
S^{1}$ is
\begin{equation}
X^{1}=l_{1}R_{1}\sigma, \quad X^{2}=l_{2}R_{2}\rho,\quad X^{I}=0,\quad
I=3,...,D,
\end{equation}
where $0\leq\sigma\leq2\pi,\: \; 0\leq\rho\leq2\pi,\: \; R_{1}$ and
$R_{2}$ are the radii of the two circles, and $l_{1}$ and $l_{2}$ are
the integers characterizing the winding numbers of the membrane around
the two circles.

In the light cone gauge, $X^{+}=p^{+}\tau$. The worldvolume metric on
this background is flat,
\begin{equation}
g_{ij}=diag(-(l_{1}l_{2}R_{1}R_{2})^{2}, (l_{1}R_{1})^{2},
(l_{2}R_{2})^{2}),
\end{equation}
and $X^{-}$ is
\begin{equation}
X^{-}=\frac{1}{2p^{+}}(l_{1}l_{2}R_{1}R_{2})^{2}\tau \;.
\end{equation}

If we consider the fluctuations $Z^{\mu}$ of the transverse coordinate
around this classical solution
\begin{equation}
X^{1}=\sigma+Z^{1},\quad X^{2}=\rho+Z^{2},\quad X^{I}=Z^{I}, \;
I=3,...,D,
\end{equation}
then, keeping only the terms of the linear order in $Z$, we find
\begin{equation}
\ddot{Z}^{1}=\partial_{\sigma}\partial_{\sigma}Z^{1} +
 \partial_{\sigma}\partial_{\rho}Z^{2},
\ddot{Z}^{2}=\partial_{\rho}\partial_{\rho}Z^{2} +
 \partial_{\sigma}\partial_{\rho}Z^{1},
\ddot{Z}^{I}=\partial_{\sigma}\partial_{\sigma}Z^{I} +
 \partial_{\rho}\partial_{\rho}Z^{I},
\end{equation}

We may fix the remaining gauge invariance. The gauge choice
$g_{0\alpha}=0$ can be solved for $\partial_{a}X^{-}$. Upon
linearization on our background, this constraint gives
\begin{equation}
\partial_{\rho}\dot{Z} ^{1}=\partial_{\sigma}\dot{Z} ^{2},
\end{equation}
from which follows the possibility
\begin{equation}
\partial_{\rho}{Z}^{1}=\partial_{\sigma}{Z}^{2}.
\end{equation}

This allows us to rewrite (24) in the form of the standard wave
equations:
\begin{eqnarray}
\ddot{Z} ^{1} = \partial_{\sigma}\partial_{\sigma}Z^{1}+
\partial_{\rho}\partial_{\rho}Z^{1}, \quad
\ddot{Z} ^{2} = \partial_{\sigma}\partial_{\sigma}Z^{2}+
\partial_{\rho}\partial_{\rho}Z^{2},\\
\ddot{Z} ^{I} = \partial_{\sigma}\partial_{\sigma}Z^{I}+
\partial_{\rho}\partial_{\rho}Z^{I}.
\end{eqnarray}

Equations of motion (28) and (29) are a special case of the equations
(20). But here it should be noted that, as follows from (22) and (26),
there is a special gauge condition, in which the general equation (20)
turns into the ordinary wave equation.

The way described above is the investigation of small variations
considering the classical solution. We may as well try to investigate
the original action (13).

Let us introduce new variables $\bar{X} ^{\mu}$:
\begin{equation}
\partial^{\alpha}\bar{X} ^{\mu}=\sqrt{|h|}h^{\alpha \beta}
 \partial_{\beta}X^{\mu}\;.
\end{equation}

This means that
\begin{equation}
\bar{h}=det(\partial_{\alpha}\bar{X} ^{\mu} \partial_{\beta}
\bar{X}_{\mu}) = sign(h)|h|^{(p+1)^{2}+1}.
\end{equation}

With these variables, the equation of motion (14) turns into the wave
equation
\begin{equation}
\partial_{\alpha}\partial^{\alpha}\bar{X} ^{\mu}=0\;,
\end{equation}
and the conditions of the constrains (17) turn into
\begin{equation}
\bar{P} ^{2} + T^{2}|\bar{h} | ^{-\frac{p^{2}}{(p+1)^{2}+1}}
 det(\partial_{i}\bar{X} ^{\mu}\partial_{j}\bar{X} _{\mu})=0,
\end{equation}
where $\bar{P} ^{\mu} \equiv \dot{\bar{X}}$ and $i,j=1,...,p$ are space indexes
 of the membrane.

For the sake of convenience, the space parameter of the membrane $\xi
_{i}\in [\xi _{i}^{a};\xi _{i}^{b}]$ is considered $\sigma_{i}\in
[0;\pi]$ and for the open dimension
\begin{equation}
X^{\mu}(\tau ,...,\sigma_{i}=0,...) \neq X
^{\mu}(\tau,...,\sigma_{i}=\pi,...),
\end{equation}
unlike for the closed dimension, where the condition of periodicity is
observed:
\begin{equation}
X^{\mu}(\tau,...,\sigma_{i},...)=X^{\mu}(\tau,...,\sigma_{i}+\pi,...)\;,
\end{equation}
or
\begin{equation}
X^{\mu}(\tau,...,\sigma_{i},...)=X^{\mu}(\tau,...,\sigma_{i}+2\pi,...)\;,
\end{equation}
depending on the spheric or toroidal type of compactification.

The border conditions for the $p$-brane in bar variables $\bar{X}^{\mu}$
 are the same like ordinary variables $X^{\mu}$. If we can the motion of the
 $p$-brane express in $X^{\mu}$ variables with the equation of motion (32),
 then the solution of this equation may be written.

In the general case, for the membrane with an arbitrary topology, when
there are $p_{0}$ open dimensions, $p_{1}$ closed dimensions with the
period $\pi$, and $p_{2}$ closed dimensions with the period $2\pi\;
(p_{0}+p_{1}+p_{2}=p)$, the solution of equation (32) may be as follows:
\begin{eqnarray}
X^{\mu}(\xi)=X^{\mu}+\frac{1}{\pi^{p}T}p^{\mu}\tau +\nonumber\\
+i\sqrt{\frac{2^{p-1}}{\pi^{p}T}}
\sum _{\bf n}n^{-1}(\alpha^{\mu}_{\bf n}e^{-in\tau}-\-
\alpha^{*\mu}_{\bf n}e^{in\tau})
\prod _{i=1}^{p}\cos n_{i} \sigma_{i} +\nonumber\\
+i\sqrt{\frac{2^{p-1}}{\pi^{p}T}}
\sum  _{\bf m}m^{-1}[(\alpha^{\mu}_{\bf m}e^{-2im\tau}-
\alpha^{*\mu}_{\bf m}e^{2im\tau})e^{-2i\bar{m} \bar{\sigma}} +\\
 + (\beta^{\mu}_{\bf m}e^{-2im\tau} - \beta^{*\mu}_{\bf m}e^{2im\tau})
e^{2i\bar{m}\bar{\sigma}}]+  \nonumber\\  +i\sqrt{\frac{2^{p-1}}
{\pi ^{p}T}}
\sum _{\bf k}k^{-1} [(\alpha^{\mu}_{\bf k}e^{-ik\tau}-\alpha^{*\mu}_{\bf k}
e^{ik\tau})e^{-i\bar{k}
\bar{\sigma}} +\nonumber\\+(\beta^{\mu}_{\bf k}e^{-ik\tau} -
 \-\beta^{*\mu}_{\bf k}
e^{ik\tau})e^{i\bar{k}\bar{\sigma}}],\quad \nonumber
\end{eqnarray}
where $X^{\mu}$ are the initial coordinates of the mass centrum and
$p^{\mu}$ is the impuls of the mass centrum of the membrane at
\begin{eqnarray}
{\bf n}\in {\bf N}^{p_{0}}\backslash 0,\quad
n=\sqrt{n_{1}^{2}+...+n_{p_{0}}^{2}}\; ;\nonumber\\ {\bf m}\in {\bf
N}^{p_{1}}\backslash 0,\;\quad
m=\sqrt{m_{1}^{2}+...+m_{p_{1}}^{2}}\; ;\nonumber\\ {\bf k}\in {\bf
N}^{p_{1}}\backslash 0,\;\quad k=\sqrt{k_{1}^{2}+...+k_{p_{2}}^{2}}\; ; \\
\bar{m}\bar{\sigma }\equiv m_{p_{0}+1}\sigma _{p_{0}+1}+...+
m_{p_{0}+p_{1}}\sigma _{p_{0}+p_{1}}\; ,\nonumber\\
\bar{k}\bar{\sigma }\equiv k_{p_{0}+p_{1}+1}\sigma _{p_{0}+p_{1}+1}
+...+ k_{p}\sigma _{p}\; .\nonumber
\end{eqnarray}
\vspace{30pt}
\section{Quantization of the model}

To investigate the quantum properties of the $p$-brane we would like to
have at our disposal the appropriate classical properties of the
original $p$-brane. The motion of the $p$-brane in the $\bar{X} ^{\mu}$
variables is the same as described by the original action (13), where
all difficulties are hidden in the constraint conditions (33). Finding
the solution of the wave equation obeying these constraint conditions is
an intricate task in itself, and its solution is yet unknown. As a first
step, let us consider the quadratic action under $X^{\mu}$ variables,
which may be interpreted as an action in the original variables
$X^{\mu}$:
\begin{equation}
S=-\frac{T}{2}\int d^{p+1}\xi h^{\alpha
\beta}\partial_{\alpha}X^{\mu}\partial_{\beta}X^{\nu}g_{\mu \nu},
\end{equation}
where $h^{\alpha \beta}=\eta^{\alpha \beta}, \; \alpha,\beta=0,...,p;\;
g_{\mu \nu}=\eta_{\mu \nu}, \; \mu,\nu=0,...,D-1$.

The action (39) is invariant respecting the global $D$-dimensional
Poincar\'{e} transformations, but not invariant under local conformal
and reparametrization transformations.

The absence of reparametrizations means the absence of the
constraints. This allows an easy quantization of the quadratic action.

Consider $X^{\mu}(\xi)$ the open $p$-brane. Then the solution of the
equation of motion (32) is like that of (37), and the density of the
energy-momentum tensor
\begin{eqnarray}
P^{\mu}_{\tau}=-{\partial {\cal L} \over \partial \dot{X}_{\mu }}=\qquad
\qquad \qquad \nonumber\\
=-i\sqrt{\frac{2^{p-1}T}{\pi^{p}}}\sum_{\bf n}(\alpha^{\mu}_{\bf
n}e^{-in\tau} - \alpha^{*\mu}_{\bf
n}e^{in\tau})\prod_{i=1}^{p}cosn_{i}\sigma_{i},
\\
\alpha ^{\mu}_{0}=\frac{1}{\sqrt {2^{p+1}\pi ^{p}T}}p^{\mu}, \quad
\quad {\bf n} \in {\bf N}^{p}. \nonumber
\end{eqnarray}

In the ligt-cone coordinates with assumption that tangent components
 $\alpha^{\pm}_{\bf n}$ are physical meaningless, like in the string case, we
have
from the commutation relations
\begin{equation}
[X^{\mu}(\tau,\sigma) , P^{\nu}_{\tau}(\tau,\sigma')]=i\eta^{\mu
\nu}\delta(\sigma - \sigma')
\end{equation}
on the quantum level
\begin{equation}
[\alpha^{i}_{\bf m} , \alpha^{j}_{\bf n}]= n \eta^{i
j}\delta_{\bf m,\bf n}\quad ,
\end{equation}
where $\alpha^{*\nu}_{\bf n}\rightarrow\alpha^{+\nu}_{\bf n}\;.$

The quantum Hamiltonian $H=\int_{0}^{\pi} d^{p}\sigma
(P^{\mu}_{\tau} \dot{X} _{\mu}-\cal L)$ is
\begin{eqnarray}
H=\frac{T}{2}\int_{0}^{\pi}(\dot{X} ^{2}+X_{1}^{2}+...+
X_{p}^{2})\- d^{p}\sigma =\nonumber\\ =\alpha^{2}_{0}+\sum _{\bf {n}}
\alpha^{+}_{\bf n}\alpha_{\bf n} + \frac{D-p-1}{2}\sum  _{\bf {n}}n,
\quad  {\bf n}\in {\bf N}^{p}\backslash 0.
\end{eqnarray}

As could be expected, the excitations of the linearized model are an
ordinary sum of the infinite number of harmonic oscillations described
by creating and annihilating operators.

The zero-point energy of the infinite number oscillators (the Casimir
energy) diverges, and for correct definition it must be regularized.

Consider the regularization by the contracted Riemann zeta-function:
\begin{equation}
{\zeta}^{'}_{p}(s)=
\sum _{\bf n}(n^{2}_{1}+n^{2}_{2}+...+n^{2}_{p})^{-s}, \quad
\quad {\bf n}_{i}\in {\bf N}^p\backslash 0,
\end{equation}
for which the following properties are known :
\begin{equation}
{\zeta}^{'}_{p}(s)=\frac{\pi^{p}}{\Gamma(s)}\int_{0}^{\infty}
dtt^{s-1}\sum
_{\bf n} exp[-\pi(n^{2}_{1}+n^{2}_{2}+...+n^{2}_{p})t]
\end{equation}
and
\begin{equation}
{\zeta}^{'}_{p}(s)=\pi^{2s-p/2}\frac{\Gamma(-s+p/2)}{\Gamma(s)}
\zeta_{p}(-s+p/2)
\quad .
\end{equation}
In our case $s=-\frac{1}{2}$. According to the definition and the
above-mentioned properties, we can find the first meanings of the
${\zeta}^{'}_{p}(-\frac{1}{2})$:
\[\begin{array}{|c|c|c|} \hline &&\\
p & {\zeta }^{'}_{p}(-\frac{1}{2}) & D_{cr}\\ &&\\
\hline
2 & 0.026 & 79.623 \\ 3 & 0.053 & 42.080 \\ 4 & 0.048 & 46.610 \\
 5 &
0.036 & 61.603 \\ 6 & 0.249 & 15.032 \\ 7 & 0.017 & 128.829 \\ 8 &
 0.011
& 199.398 \\
\hline
\end{array}\]

Then, substituting the quantities ${\zeta}^{'}_{p}(-\frac{1}{2})$ in
(43), we obtain the undiverging meanings of the Casimir energy and,
correspondingly, good properties of the Hamiltonian $H$.

We remember that in the quantum case we have no constraints for this
model. But we may impose "by hand" an additional condition
$H|\phi\rangle=0$. In this case, we obtain that for the existence of a
massless vector, the coefficients at the second term in (43) must equal
to one. This condition gives $D=D_{cr}=1+p+ 2(\sum_{n>0} n)^{-1}$.
 Hence, the ground state of this model is a tachyon.

Now, let us consider $X^{\mu }(\xi )$ as a closed $p$-brane with the
period $2\pi $ (a toroidal type of the $p$-brane). Then, the solution
 of
the equations of motion is like that of (37), and the density of the
energy-momentum tensor
\begin{eqnarray}
P^{\mu}_{\tau}=-{\partial {\cal L} \over \partial \dot{X}_{\mu }} =
\qquad
\qquad \qquad \nonumber\\ =-i\sqrt{\frac{2^{p-1}T}{\pi ^{p}}}
\sum _{\bf k} [(\alpha^{\mu}_{\bf k}e^{-ik\tau}-\alpha^{*\mu}_{\bf k}
e^{ik\tau})
e^{-i\bar{k}
\bar{\sigma}} +\\
+(\beta^{\mu}_{\bf k}e^{-ik\tau} -
\-\beta^{*\mu}_{\bf k}e^{ik\tau})e^{i\bar{k}\bar{\sigma}}],\nonumber\\
\alpha ^{\mu}_{0}=\beta ^{\mu}_{0}\frac{1}{2\sqrt {2^{p+1}\pi ^{p}T}}
p^{\mu}, \quad \quad  {\bf k} \in {\bf N}^{p}. \nonumber
\end{eqnarray}
The left-right symmetry condition gives us the correlation between
 the
coefficients $\alpha ^{\mu}_{\bf k}$ and $\beta ^{\nu}_{\bf k}$
\begin{equation}
\beta ^{\mu}_{\bf k}=\alpha ^{\mu}_{-\bf k}\; .
\end{equation}

In this case, from the commutation relations (41) it follows that
\begin{eqnarray}
[\alpha^{\mu}_{\bf m} , \alpha^{+\nu}_{\bf n}]= n \eta^{\mu
\nu}\delta_{\bf m,\bf n}\quad ,\quad [\beta^{\mu}_{\bf m} ,
\beta^{+\nu}_{\bf n}]= n \eta^{\mu \nu}\delta_{\bf m,\bf n}
\quad ,\\
{[\alpha^{\mu}_{\bf m} , \beta^{+\nu}_{\bf n}]}=
 {[\alpha^{\mu}_{\bf m}
, \beta^{+\nu}_{\bf n}]}=0\;.\quad \qquad \qquad
\end{eqnarray}
The quantum Hamiltonian
\begin{eqnarray}
H=\frac{T}{2}\int_{0}^{2\pi}(\dot{X} ^{2}+X_{1}^{2}+...+
X_{p}^{2})\- d^{p}\sigma =\nonumber\\ \alpha^{2}_{0}+\beta^{2}_{0}+
\frac{1}{2}\sum _{\bf {k}}\{ \alpha^{+}_{\bf k},\alpha_{\bf k}\}+
\frac{1}{2}\sum _{\bf {k}}\{ \beta^{+}_{\bf k},\beta_{\bf k}\}=
H _{L}+H _{R}\; ,
\end{eqnarray}
where $H _{L}(H _{R})$ depends only on $\alpha^{\mu}_{\bf
k}(\beta^{\mu}_{\bf k})$ variables and ${\bf k}\in {\bf N}^p\backslash 0$.

In the case of the toroidal $p$-brane, we have two different
possibilities: (a) to impose a more detailed condition $H
_{L}|\varphi \rangle =H _{R}|\varphi \rangle =0$ or an equivalent
$H |\varphi \rangle =H _{L}|\varphi \rangle =0 \quad (H
_{L}|\varphi \rangle =H _{R}|\varphi \rangle =0)$; (b) using the
discrete symmetry condition $X^{\mu}(\sigma ,\tau )=X^{\mu}(-\sigma
,\tau )$ and, consequently, the correlation between $\alpha^{\mu}_{\bf
k}$ and $\beta^{\nu}_{\bf k}$ operators, we may impose only one
condition $H |\varphi \rangle =0$.

In the first case, we have the same properties for the closed $p$-brane
as for the open one:
\begin{eqnarray}
H_{L}=\beta^{2}_{0}+\sum _{\bf {k}} \beta^{+}_{\bf k}
\beta_{\bf k}+\frac{D-p-1}{2}\sum  _{\bf {k}}k\;, \nonumber \\
H_{L}=\alpha^{2}_{0}+\sum _{\bf {k}} \alpha^{+}_{\bf k}
\alpha_{\bf k}+\frac{D-p-1}{2}\sum  _{\bf {k}}k
\end{eqnarray}
where ${\bf k}\in {\bf N}^p\backslash 0$ and, according to the conditions (a),
 we obtain a tachyon in a ground
state and the $D_{cr}$ corresponding to that in the table for the open
$p$-brane.

In the second case we may express $H$ only in the terms of the
right (left) operators $\alpha^{\mu}_{\bf k}(\beta^{\mu}_{\bf k})$:
\begin{equation}
H_{L}=\alpha^{2}_{0}+\sum _{\bf {k\neq 0}} \alpha^{+}_{\bf k}
\alpha_{\bf k}+\frac{D-p-1}{2}\sum  _{\bf {k\neq 0}}k\quad ,
 \end{equation}
where $k=\sqrt{k_{1}^{2}+...+k_{p}^{2}}$.

Using the definition of the ordinary Riemann zeta-function \cite{Dow}
\begin{equation}
{\zeta}_{p}(s)=\sum _{k\neq 0},
(k^{2}_{1}+k^{2}_{2}+...+k^{2}_{p})^{-s},
\end{equation}
with the same properties (45),(46), we
may find the first meanings of ${\zeta}_{p}(-\frac{1}{2})$:
\[\begin{array}{|c|c|c|} \hline &&\\
p & {\zeta }_{p}(-\frac{1}{2}) & D_{cr}\\ &&\\
\hline
2 & -0.229 & 11.734\\ 3 & -0.267 & 11.491\\ 4 & -0.297 & 11.734\\ 5 &
-0.325 & 12.154\\ 6 & -0.373 & 13.362\\ 7 & -0.407 & 12.914\\ 8 & -0.462 &
13.329\\
\hline
\end{array}\]

Then, substituting the quantities ${\zeta }_{p}(-\frac{1}{2})$ in (53 ),
we find no divergencies of the Hamiltonian $H$. In this case, the
ground state of the toroidal $p$-brane is also a tachyon, and the
critical dimension $D_{cr}=1+p-2(\sum_{k\neq 0}k)^{-1}$.

\section{A supersymmetric linearized model}

It is of interest to examine the supersymmetric case of the bosonic
$p$-brane in the GS and NSR approaches. Let us consider the
supersymmetric linearized model in the NSR approach. Let $p=2$. Passing
over to the $p\geq2$ will be simple.

The direct generalization of the linearized model of the bosonic action
is

\begin{equation}
S=-\frac{T}{2}\int d^{3}\xi(\partial_{\alpha}
X^{\mu}\partial^{\alpha}X_{\mu}+i\bar {\psi }^{\mu}
\gamma^{\beta}\partial_{\beta}\psi_{\mu}),
\end{equation}
where $\psi^{\mu}$ is the Majorana spin-vector, $\{
\gamma^{\alpha},\gamma^{\beta}\} =-2\eta^{\alpha \beta}$. We shall use
the basis for $\gamma^{\alpha}$:
\begin{equation}
\gamma^{0}=\left(\begin{array}{cc} 0 & {-i}\\i & {0}\end{array}\right)
,\quad
\gamma^{1}=\left(\begin{array}{cc} 0 & {i}\\i & {0}\end{array}\right)
,\quad
\gamma^{2}=\left(\begin{array}{cc} i & {0}\\0 & {-i}\end{array}\right).
\end{equation}

This action is invariant under the transformations
\begin{equation}
\delta X^{\mu}=i\bar{\eta} \psi ^{\mu},\quad \quad
\delta \psi^{\mu}=(\gamma ^{\alpha}\partial_{\alpha}X^{\mu})\eta\quad .
\end{equation}

The equations of motion, which follow from the action of the super
$p$-brane, are
\begin{equation}
\partial _{\alpha }\partial ^{\alpha }X^{\mu}=0, \quad \partial_{2}
\psi^{\mu}_{1} + (\partial_{1}-\partial_{0})\psi^{\mu}_{2} = 0,\quad
(\partial_{1} + \partial_{0})\psi^{\mu}_{1} - \partial_{2}
\psi^{\mu}_{2}
= 0.
\end{equation}

For the variables $X^{\mu}$ we may use the same solution as in the
bosonic case (37).  Let the solutions of the equations of motion for the
fermionic part have the following form:
\begin{eqnarray}
\psi^{\mu}_{1}(\xi )=\sum  _{\bf n}d^{\mu(1)}_{\bf n}
 e^{-i(n_{0}\tau + n_{1}\sigma_{1} + n_{2}\sigma_{2})},\\
\psi^{\mu}_{2}(\xi )= \sum _{\bf n}d^{\mu (2)}_{\bf n}
 e^{-i(n_{0}\tau - n_{1}\sigma_{1} + n_{2}\sigma_{2})},
\end{eqnarray}
where ${\bf n}\in {\bf Z}^3$

In this case the equation of motion imposes restrictions on the
coefficients $d^{\mu (1)}_{\bf n}$ and $d^{\mu
(2)}_{\bf n}$:
\begin{equation}
d^{\mu (2)}_{\bf n}=n_{2}/(n_{0}-n_{1})d^{\mu
(1)}_{\bf n}=(n_{0}+n_{1})/n_{2}d^{\mu(1)}_{\bf n}
\end{equation}

The Hamiltonian of such system equals to
\begin{equation}
H = \frac{T}{2}\int d^{3}\xi [\dot{X} ^{2} + X_{1}^{2} +
 X_{2}^{2} +
\frac{i}{2}(\bar{\psi}^{\mu}\gamma^{0}\dot{\psi} _{\mu} -
 \dot{\bar{\psi} }
^{\mu}\gamma^{0}\psi_{\mu})]
\end{equation}

In the quantum case, the coefficients $d^{\mu (i)}_{\bf n}$
obey the anticommutation relations:
\begin{equation}
\{ d^{\mu (i)}_{\bf m}, d^{\nu (j)}_{\bf n}\}
=\eta^{\mu \nu}\delta_{\bf m,\bf -n}\quad .
\end{equation}

Therefore, the Hamiltonian is the sum of the bosonic and fermionic
oscillations:
\begin{equation}
H =\alpha_{0}^{2} + \sum_{\bf n} [\alpha^{+}_{\bf n}\alpha _
{\bf n}] +
\sum_{i,{\bf n}} [nd^{+(i)}_ {\bf n}d^{(i)}_{\bf n}],
\end{equation}
for which there is no Casimir energy. This is what must be the case with
the supersymmetric model.

The initial action does not contain any auxiliary metric on the
worldvolume, hence the constraints in the system are absent, too. Like
in the bosonic case, we may impose an additional condition $H =0$
and consider it in the quantum case as well.

In this case, we find that in the supersymmetric model the condition
$H=0$ gives us massless ground states and no critical dimensions
whatsoever.

\section{Discussion}

In this article we have considered the simplest case of the bosonic and
fermionic membranes, when they contain only linear terms in their
equations of motion. The general situation is much more complicated.

An essential point of our consideration is imposing additional
conditions like $H=0$. But in the case of the linearized model we
can consider these conditions as a certain remnant constraint condition
like $L_{n}=0$.

One would remark that $D_{cr}$ in the bosonic case is not an integer
and, consequently, has no physical meaning. Indeed, in all considered
cases $D_{cr}\neq \bf {N}$. But even in the case when $D_{cr}\in \bf
{N}$, $D_{cr}$ has no physical meaning. The point is that we cannot pick
out physical states among all possible states in the Hilbert space, as
we have not enough constraints or the conditions like those and can not
obtain the physical sector. On the other hand, the discrete values of
the spacetime dimension $D_{cr}$ imply the existence of the fractal
properties of the extended objects. Some of the aspects of these properties
are considered in \cite{P2}.

In the supersymmetric case we have additional possibilities to impose
condition, at which the supercurrent $J^{\alpha} = K\gamma^{\beta}
\gamma^{\alpha} \psi^{\mu} \partial_{\beta}X_{\mu}$ vanishes. In this
case the condition $J^{\alpha}=0$ is equivalent to six conditions
$\partial _{\alpha }X^{\mu }\psi _{\mu }^{i}=0$ or their Fourier
transformation $F_{\bf n}^{\alpha i}=\int _{-\pi }^{\pi }d^{2}\sigma
e^{i\vec{n} \vec{\sigma }}\partial _{\alpha }X^{\mu }\psi _{\mu
}^{i}$. The supersymmetric action contains the constraints
$F_{\bf n}^{\alpha i}=0$. We may also express this quantity in the
$\alpha_{\bf n}, d^{(i)}_ {\bf n}$ variables and consider the quantum
case, but this will be also not enough to distinguish the physical
sector. Nevertheless, due to the quadratic action we can analytically
calculate the partition function and transition amplitude for this
model.

The linearized model allows us to separate linear and nonlinear effects
in the general (super)$p$-brane. For instance, in \cite{Mar} , due to
the restriction of the constraint condition for the bosonic $p$-brane,
$D_{cr}$ has been obtained, whereas the purely linearized model has no
critical dimensions. This means that in \cite{Mar} a nontrivial
conformity between the linearized model and the imposed constraint
condition was obtained.

We may try to impose sufficient constraint conditions as an additional
condition, but in this case a very important question arises: how to
conform the solution of the equation of motion with the constraint
conditions? We can make it sure that in the bosonic sector the simplest
quadratic constraints $\dot X^{2}+X_{;1}^{2}+...+X_{;p}^{2}=0, \; \dot
X^{\mu }X_{;i\mu }=0$, which are a natural generalization of the string
constraints, cannot coexist with the solutions of the linear wave
equation of motion for the bosonic $p$-brane.  Thus, the conformity
between the solution of the equation of motion in the linearized model
and the additional constraint conditions is nontrivial and of interest
in itself.

On the other hand, we may not only use global supersymmetry and
vanishing of the supercurrent $J^\alpha,$ but also the condition of
local supersymmetry may be imposed. Indeed, we may use the linearized
model of the (super)$p$-brane with local supersymmetry and try to find
the conformity between the solutions and constraints. However, (1) it is
not clear how to do it even in a less complicated case without
supersymmetry, and (2) this will be not enough to distinguish the
physical sector, either.

Thus, we may consider the linearized model an auxiliary model of the
(super)$p$-brane. An important aspect of this consideration is the
possibility to separate the physical properties belonging to the
linearized model from other properties characteristic of the essentially
nonlinear behavior of the relativistic (super)$p$-brane.

\section{Acknowledgements}

Author wants to express his gratitude to A.Bytsenko, S.Odintsov,
 R.Zaikov for the useful comments and the Swedish Institute for Grant
304/01 GH/\-MLH, which gave him the possibility to enjoy the kind
hospitality of Prof. Antti Niemi, Doc. Staffan Yngve and all members of
the Institute of Theoretical Physics, Uppsala University.

\end{document}